\documentclass[prd,aps,twocolumn,a4paper,floatfix,showpacs,nofootinbib]{revtex4-1}

\usepackage[utf8]{inputenc}  
\usepackage[T1]{fontenc}
\usepackage{graphicx,psfrag}
\usepackage{mathrsfs}
\usepackage{amsmath,amsfonts,amssymb,pifont}
\usepackage{multirow,enumerate}
\usepackage{comment,hyperref}
\usepackage{color}
\usepackage{acronym}
\usepackage{xspace}
\usepackage[normalem]{ulem}
\usepackage{mathtools}
\usepackage{subfigure}
\usepackage{makecell}

\newacro{BH}{black hole}
\newacro{NS}{neutron star}
\newacro{PN}{Post-Newtonian}
\newacro{BBH}{binary black hole}
\newacro{BNS}{binary neutron star}
\newacro{EOB}{effective-one-body}
\newacro{NR}{numerical relativity}
\newacro{GW}{gravitational wave}
\newacro{EOS}{equation-of-state}

\newcommand{\be}{\begin{equation}}
\newcommand{\ee}{\end{equation}}
\newcommand{\bea}{\begin{eqnarray}}
\newcommand{\eea}{\end{eqnarray}}
\newcommand{\bel}{\begin{align}}
\newcommand{\eel}{\end{align}}

\newcommand{\cmark}{\ding{51}}%
\newcommand{\xmark}{\ding{55}}%
\newcommand{\linf}{\texttt{LALInference}}

\newcommand{\imrpnrt}{\texttt{IMRPNRT}}
\newcommand{\imrdnrt}{\texttt{IMRDNRT}}
\newcommand{\tfnrt}{\texttt{TF2NRT}}
\newcommand{\seobnrt}{\texttt{SEOBNRT}}

\newcommand{\imrppnt}{\texttt{IMRPPNT}}
\newcommand{\tfpnt}{\texttt{TF2PNT}}

\newcommand{\tf}{\texttt{TF2}}
\newcommand{\seob}{\texttt{SEOB}}

\def\Msun{{\rm M_{\odot}}}
\def\GMc2{{\rm G M_{\odot} c^{-2}}}

\def\ct{\tilde{c}}

\def\nt{\tilde{n}}
\def\dt{\tilde{d}}
\def\NRtidal{\texttt{NRTidal}\xspace}

\def\SEOBNRROM{\texttt{SEOBNRv4\_ROM}\xspace}
\def\SEOBNRROMNRtidal{\texttt{SEOBNRv4\_ROM\_NRTidal}\xspace}
\def\PhenomDNRtidal{\texttt{IMRPhenomD\_NRTidal}\xspace}
\def\PhenomPNRtidal{\texttt{IMRPhenomPv2\_NRTidal}\xspace}
\def\PhenomP{\texttt{IMRP}\xspace}
\def\PhenomD{\texttt{IMRD}\xspace}
\def\TEOBResumS{\texttt{TEOBResumS}\xspace}
\def\SEOBNRv4T{\texttt{SEOBNRv4T}\xspace}

\def\TaylorF{\texttt{TaylorF2}\xspace}

\usepackage{color}

\definecolor{cyan}{rgb}{0,0.9,0.9}
\definecolor{orange}{rgb}{0.9,0.5,0}
\definecolor{magenta}{rgb}{1,0,1}
\definecolor{purple}{rgb}{0.8,0.4,0.8}
\definecolor{gray}{rgb}{0.5,0.5,0.5}
\definecolor{mygreen}{rgb}{0.1,0.8,0.1}
\definecolor{darkblue}{rgb}{0.0,0.0,0.6}

\begin{document}

\title{Extracting source properties from binary neutron star gravitational wave signals: 
effects of the point-particle baseline and tidal description} 

\title{Waveform systematics for binary neutron star gravitational wave signals: 
effects of the point-particle baseline and tidal descriptions} 

\author{Anuradha Samajdar$^1$}
\author{Tim \surname{Dietrich}$^1$}

\affiliation{${}^1$ Nikhef, Science Park, 1098 XG Amsterdam, The Netherlands}

\date{\today}

\begin{abstract}
Gravitational wave (GW) astronomy has consolidated its role as a new observational 
window to reveal the properties of compact binaries in the Universe. In particular, the  
discovery of the first binary neutron star coalescence, GW170817, led to 
a number of scientific breakthroughs as the possibility to place 
constraints on the equation of state of cold matter at supranuclear densities. 
These constraints and all scientific results based on them require accurate 
models describing the GW signal to extract the source properties from the measured signal. 

In this article, we study potential systematic biases during the extraction 
of source parameters using different descriptions for both, the 
point-particle dynamics and tidal effects.
We find that for the considered cases the mass and spin recovery 
show almost no systematic bias with respect to the chosen waveform model. 
However, the extracted tidal effects can be strongly 
biased, where we find generally that 
Post-Newtonian approximants predict neutron stars with 
larger deformability and radii than numerical relativity tuned models. 
Noteworthy, an increase in the Post-Newtonian order in the tidal phasing 
does not lead to a monotonic change in the estimated properties.

We find that for a signal with strength similar to GW170817, but observed with design sensitivity, 
the estimated tidal parameters can differ by more than a factor of two
depending on the employed tidal description of the waveform approximant.
This shows the current need for the development of better waveform models to 
extract reliably the source properties from upcoming GW detections. 
\end{abstract}

\maketitle

\section{Introduction}

With GW170817, the first detected gravitational wave (GW) signal 
emitted from a binary neutron star (BNS) coalescence, 
the LIGO-Virgo Collaboration (LVC) managed to place 
constraints on the unknown equation of state (EOS)
for supranuclear dense matter~\cite{TheLIGOScientific:2017qsa}. 
Initiated by this study and results obtained by 
the source properties extracted from the GW signal 
a number of groups were able to obtain even tighter bounds 
on the NS radii and EOS either by the combination 
of the information obtained from the GWs with 
electromagnetic observations~\cite{Radice:2017lry,Bauswein:2017vtn,Coughlin:2018miv},
or by statistical methods based on a large set of possible EOSs 
combining GW information with new insights from nuclear 
physics theory~\cite{Annala:2017llu,Most:2018hfd}. 
Also the LVC updated their first analysis~\cite{TheLIGOScientific:2017qsa} 
incorporating the known source location, 
employing updated waveform approximants, 
using re-calibrated Virgo data, and 
starting at a lower frequency threshold. 
The improved analysis, Refs.~\cite{Abbott:2018wiz,Abbott:2018exr}, 
determined the NS radii to be $11.9^{+1.4}_{-1.4}\rm km$. 
In addition to the LVC, De et al.~\cite{De:2018uhw} and 
Dai et al.~\cite{Dai:2018dca} re-analyzed GW170817 
with slightly different methods finding overall consistency. 

Just from this single detection, it was already possible to 
show that several proposed nuclear physics EOSs can hardly explain the 
observed signal~\cite{TheLIGOScientific:2017qsa,Abbott:2018wiz,Abbott:2018exr}.
The increasing sensitivity of advanced GW detectors
over the next years will lead to a growing number of detections 
of merging BNSs in the near future~\cite{Abbott:2016ymx} allowing 
an even more precise measurement of the supranuclear EOSs. \\

Extracting the source properties of a compact binary 
from GW data requires Bayesian analysis involving 
calculation of a multi-dimensional likelihood function of the data 
with an expected waveform model. 
In context of the GW data analysis considered in this article, 
this is done using the Bayesian inference module
\linf{}~\cite{Veitch:2014wba}. 
Consequently, the generation of individual waveforms needs 
to be efficient to allow evaluation of a large number of templates, 
but also needs to be accurate enough for a precise measurement 
of the intrinsic source parameters.

Over the last years, there has been significant progress in the development of 
BNS waveform approximants capturing the strong-field and
tidally dominated regime from the early inspiral up to the merger of the two stars.
Exemplary cases for such developments are the improved analytical post-Newtonian (PN) 
based waveform models, e.g.~\cite{Damour:2012yf,Agathos:2015uaa}, the existing 
state-of-the-art tidal waveform models in the time domain as discussed in 
\cite{Bernuzzi:2014owa, Hotokezaka:2016bzh, Hinderer:2016eia, 
Steinhoff:2016rfi, Dietrich:2017feu,Nagar:2018zoe} based on the
effective-one-body (EOB) description of the general-relativistic
two-body problem \cite{Buonanno:1998gg, Damour:2009wj}, or 
closed-form tidal approximants combining PN, tidal EOB, 
and numerical relativity (NR) information~\cite{Dietrich:2017aum,Dietrich:2018upm,
Dietrich:2018uni,Kawaguchi:2018gvj}. 
In particular the model of Ref.~\cite{Dietrich:2017aum} 
has proven its importance for the interpretation of 
GW170817~\cite{TheLIGOScientific:2017qsa,Abbott:2018wiz,Abbott:2018exr,Dai:2018dca} 
due to the significantly smaller 
computation costs compared to time domain tidal EOB models, 
but the better accuracy with respect 
to analytical PN models~\cite{Dietrich:2018uni}.  
Reference~\cite{Jimenez-Forteza:2018buh} studied the effects of 
higher order tidal terms in the PN-expansion in addition to 
studying the effects of magnetic tidal polarizability and tidal-spin coupling.
They concluded that the inclusion of the tidal-tail term at 6.5PN
deteriorates the PN series convergence. They also established that the magnetic tidal 
deformability has negligible effects on the phasing. This has also been 
studied exclusively in Ref.~\cite{Pani:2018inf}.

While Ref.~\cite{Dietrich:2018uni} indicated already that NR based waveform models 
allow more precise and stringent extraction of the source properties, the authors 
did not perform a full Bayesian analysis to determine the 
influence of different tidal and point-particle (PP) 
descriptions on the estimated binary parameters.  

The first Bayesian study for the estimation of tidal deformability parameters was carried out
in~\cite{DelPozzo:2013ala} which included tidal corrections up to 1PN (6PN in phase). They reported 
tens of detections required to constrain the tidal deformability parameter to $\sim$ 10\% accuracy
and distinguish between different types of EOSs by expressing the tidal deformability as a linear expansion 
in mass. 
Following this work, Ref.~\cite{Agathos:2015uaa} extended
the analysis including tidal phasing terms up to 
2.5PN and modify the termination criterion to be the 
minimum of the contact frequency~\cite{Maselli:2013mva} 
and the frequency at the last stable orbit (LSO). 
Their work also included the quadrupole-monopole 
deformation at leading order~\cite{Poisson:1997ha,Yagi:2013bca}, which 
did not affect parameter estimation (PE) for 
the considered configurations. 
In addition, the tidal deformability parameter 
was expanded to include up to a quadratic function of mass.
In~\cite{Lackey:2014fwa}, the pressure and 
the adiabatic indices were sampled directly. 
An analytical approach to study the systematic and 
statistical uncertainties arising from 
neglecting physical effects in the estimation of the NS Love number
was made by~\cite{Favata:2013rwa}, a similar analysis based on 
a Bayesian approach, was made in Ref.~\cite{Wade:2014vqa}.
Very recently,~\cite{Dudi:2018jzn} investigated NR-based tidal waveform models
and showed the importance of the inclusion of tidal effects for the extraction of the 
NS masses and spins from the GW signal for high signal-to-noise ratios.
An exhaustive PE analyses on GW170817-like simulations have 
been performed in~\cite{Abbott:2018wiz} and used the \NRtidal 
description for the first time in PE. The simulations were 
done with a tidal EOB-based model~\cite{Hinderer:2016eia} and the recovery 
was performed with aligned-spinning models. 
This work probed different mass-ratios 
for non-spinning signals all consistent with 
GW170817-supported EOS models. 

Unfortunately, none of the existing works allowed a clear distinction between possible biases 
introduced by the binary black hole (BBH) or PP description and 
the particular inclusion of tidal effects. 
To fill this gap, we study equal and unequal mass BNS signals 
and employ four different PP baselines and five different tidal descriptions. 
This allows a systematic study to understand possible biases in the 
source properties including the tidal parameters. 

The article is structured as follows. 
In Sec.~\ref{sec:models} we discuss the numerical methods and 
the employed waveform approximants. 
In Sec.~\ref{sec:results:PP} we assess the imprint of the PP 
description of the waveform model and in Sec.~\ref{sec:results:tides}
we keep the PP baseline fixed but vary the tidal description. 
We conclude in Sec.~\ref{sec:conclusion}.

\section{Numerical settings and waveform models}
\label{sec:models}

To determine systematic biases on the extraction of 
source properties from a detected BNS signal, we simulate
a number of waveforms and recover these with waveforms using 
different combinations of PP baselines and tidal descriptions.
New waveform approximants constructed for this article 
are the \texttt{IMRPhenomPv2\_PNTidal}, a combination of \PhenomP 
and PN based tidal phasing given by Eq.~\ref{eq:PsiPNT} to 
\texttt{IMRPhenomPv2} (henceforth \PhenomP), 
and \texttt{TaylorF2\_NRTidal} (henceforth \tfnrt), 
which is a combination of the \NRtidal approximation 
with the \TaylorF (henceforth \tf) model.
Furthermore, we consider the existing PP models 
\texttt{IMRPhenomD} (\PhenomD) and \SEOBNRROM (\seob).
Throughout this article, we refer to \tf\ extended with PN-tides 
as \tfpnt, to \PhenomPNRtidal as \imrpnrt, 
to \PhenomDNRtidal as \imrdnrt, 
and to \SEOBNRROMNRtidal as \seobnrt. 
Details about the BBH baselines 
and the tidal descriptions are given below.

\subsection{Waveform models}

Our work is based on four different frequency domain BBH 
waveform models, summarized in Tab.~\ref{tab:waveform_models}. 
These baseline models are then extended with different tidal descriptions. 
All chosen models are fast enough to be used 
for parameter estimation studies using \linf{} and  
have been employed for the analysis of GW170817 presented in 
Ref.~\cite{Abbott:2018wiz}. 

The frequency domain waveform is given by 
\begin{equation}
 \tilde{h}(f) = \tilde{A}(f) e^{-i \Psi(f)},
\end{equation}
where the frequency domain phase $\Psi(f)$ can be approximated as 
a sum of the non-spinning PP contribution, 
a spin-orbit (SO) contribution, a spin-spin (SS) 
contribution, and tidal contributions (Tides), i.e., 
\begin{equation}
 \Psi(f) = \Psi_{\rm PP}(f) + \Psi_{\rm SO}(f) + \Psi_{\rm SS}(f) +  \Psi_{\rm Tides}(f). 
\end{equation}

\subsubsection{BBH-baseline models}

\begin{table*}
\begin{footnotesize}
\begin{tabular}{l|c|c|c|c|c|c|c}
Model name & point-particle & spin-orbit & spin-spin  & Quadrupole-Monopole   &  Precession \\
\hline
 \TaylorF(\tf)     & 
 3.5PN~\cite{Sathyaprakash:1991mt} &
 3.5PN~\cite{Bohe:2013cla,Bohe:2013cla}          & 
 3PN~\cite{PhysRevD.79.104023,PhysRevD.71.124043,Bohe:2015ana,Mishra:2016whh} & 
 \xmark \quad 3PN for BBHs &   
 \xmark \\
 \SEOBNRROM(\seob)     & 
 \multicolumn{3}{c|}{EOB model with NR calibration~\cite{Bohe:2016gbl,Puerrer:2014fza}} &
 \xmark \quad as BBH &   
 \xmark \\
 \texttt{IMRPhenomD}(\PhenomD)     & 
 \multicolumn{3}{c|}{phenom.~model calibrated to \tf\ and EOB/NR-hybrids~\cite{Husa:2015iqa,Khan:2015jqa,Hannam:2013oca}}  &
 \xmark \quad as BBH &   
 \xmark \\
 \texttt{IMRPhenomPv2}(\PhenomP)     & 
 \multicolumn{3}{c|}{same spin-aligned model as \PhenomD} &  3PN~EOS dependent\cite{Poisson:1997ha,YAGI20171} &   
 \cmark~\cite{Schmidt:2012rh,Schmidt:2014iyl} \\
\end{tabular}
\end{footnotesize}
\caption{Overview about the employed baseline waveform models used in this article. }
\label{tab:waveform_models}

\end{table*}
Let us very briefly review the main characteristics of the underlying BBH models.
The \tf\ model is a pure analytical PN-based approximant including 
PP and aligned SO terms to 3.5PN order and SS effects 
up to 3PN~\cite{Sathyaprakash:1991mt,Blanchet:1995ez,Damour:2001bu,Blanchet:2004ek,
Blanchet:2005tk, Bohe:2013cla, Blanchet:2013haa,Arun:2008kb, 
Mikoczi:2005dn, Bohe:2015ana, Mishra:2016whh}.

The \seob\ approximant is a frequency domain reduced order model 
constructed following Ref.~\cite{Puerrer:2014fza} and it is constructed 
from the aligned-spin EOB model presented in~\cite{Bohe:2016gbl}.

\PhenomD is based on the aligned-spin model presented in Refs.~\cite{Husa:2015iqa,Khan:2015jqa} 
and calibrated to untuned EOB waveforms of~\cite{Taracchini:2013rva} and NR hybrids~\cite{Husa:2015iqa,Khan:2015jqa}. 
Finally, \PhenomP~\cite{Hannam:2013oca} uses \PhenomD as an underlying spin-aligned approximant which is then 
`twisted up' to include precession effects as described in Refs.~\cite{Schmidt:2012rh,Schmidt:2014iyl}. 

\subsubsection{Inclusion of tidal effects}

To incorporate tidal effects and allow for a proper modeling of BNS waveforms, 
we employ two different approaches by either including PN-based 
tidal approximations~\cite{Vines:2011ud,Damour:2012yf} or including an \NRtidal 
approximation~\cite{Dietrich:2017aum,Dietrich:2018upm,Dietrich:2018uni} which is calibrated 
in the high-frequency region to the analytic EOB model 
of~\cite{Bernuzzi:2014owa,Nagar:2018zoe} 
and NR simulations~\cite{Dietrich:2017aum,Dietrich:2018upm,Dietrich:2018phi}.

Independent of the exact form of the tidal phase contribution, 
tidal effects enter the waveform's phase due to the tidal deformabilities of the individual 
components in the binary via
\begin{equation}
 \Lambda_{A,B} = \frac{2}{3} \frac{k_2^{A,B}}{C_{A,B}^5},
\end{equation}
where $k_2^{A,B}$ denote the Love numbers of the individual stars 
describing the static quadrupolar deformation of one body 
in the gravitoelectric field of the companion~\cite{Damour:2009vw}, 
and $C_{A,B}=M_{A,B}/R_{A,B}$ denote the compactness of the individual stars.
In terms of the characteristic PN-parameter $x = x(f) = (\pi M f)^{2/3}$,
the PN tidal phase takes the form 
\begin{widetext}
\begin{align}
 \Psi_{\rm PNT}
 & = \frac{3}{128 \eta x^{5/2}} \Lambda_A X_A^4 
  \left( \underbrace{-24 (12-11X_A) x^{5}}_{\rm 5PN} + 
 \underbrace{\frac{5}{28} (3179-919 X_A -2286 X_A^2 +260 X^3_A) x^{6}}_{\rm 6PN} + 
 \underbrace{24 \pi (12-11 X_A) x^{6.5} }_{\rm 6.5PN}
 \right. \nonumber  \\
 & \underbrace{ - 24 \left( \frac{39927845}{508032} - \frac{480043345}{9144576}X_A + \frac{9860575}{127008}X_A^2 - 
 \frac{421821905}{2286144}X_A^3 + \frac{4359700}{35721}X_A^4 - \frac{10578445}{285768}X_A^5 \right) x^{7}}_{\rm 7PN^*} 
 \nonumber \\
 & \left. + \underbrace{ \frac{\pi}{28} (27719 - 22127 X_A + 7022 X_A^2 -10232 X_A^3) x^{7.5}}_{\rm 7.5 PN} \right)  
 + [A \leftrightarrow B], \label{eq:PsiPNT}
\end{align}
\end{widetext}
where PN orders marked with $^*$ are incomplete and contain yet unknown parameters which for the analysis of
this paper have been set to zero and according to~\cite{Damour:2012yf} might be negligible. 
$X_A$ and $X_B$ denote respectively the mass fractions $M_A/M$ and $M_B/M$, $M=M_A + M_B$ being the total mass 
of the binary.

For the analyses done with waveform models including the \tf\ PP phasing, we
have truncated the waveform at a frequency which is the minimum of ISCO or the contact frequency. 
The latter is given by
\begin{equation}
f_{\mathrm{contact}} = \frac{1}{\pi} \left(\frac{M}{(R(M_A) + R(M_B))^3}\right)^{1/2},
 \label{eqn:fcont}
\end{equation}
where the radii of the component masses are determined using the phenomenological
relation between the compactness $C$ and the tidal deformability $\Lambda$ by~\cite{Maselli:2013mva}
\begin{align}
C_{A,B} = & 0.371 - 3.91\times10^{-2}\ln \Lambda_{A,B} + \nonumber \\ 
        & +  1.056\times10^{-3}(\ln\Lambda_{A,B})^2.
 \label{eqn:loveCompact}
\end{align}
\begin{figure*}[t]
 \mbox{\subfigure{ 
   \includegraphics[keepaspectratio,width=0.48\textwidth]{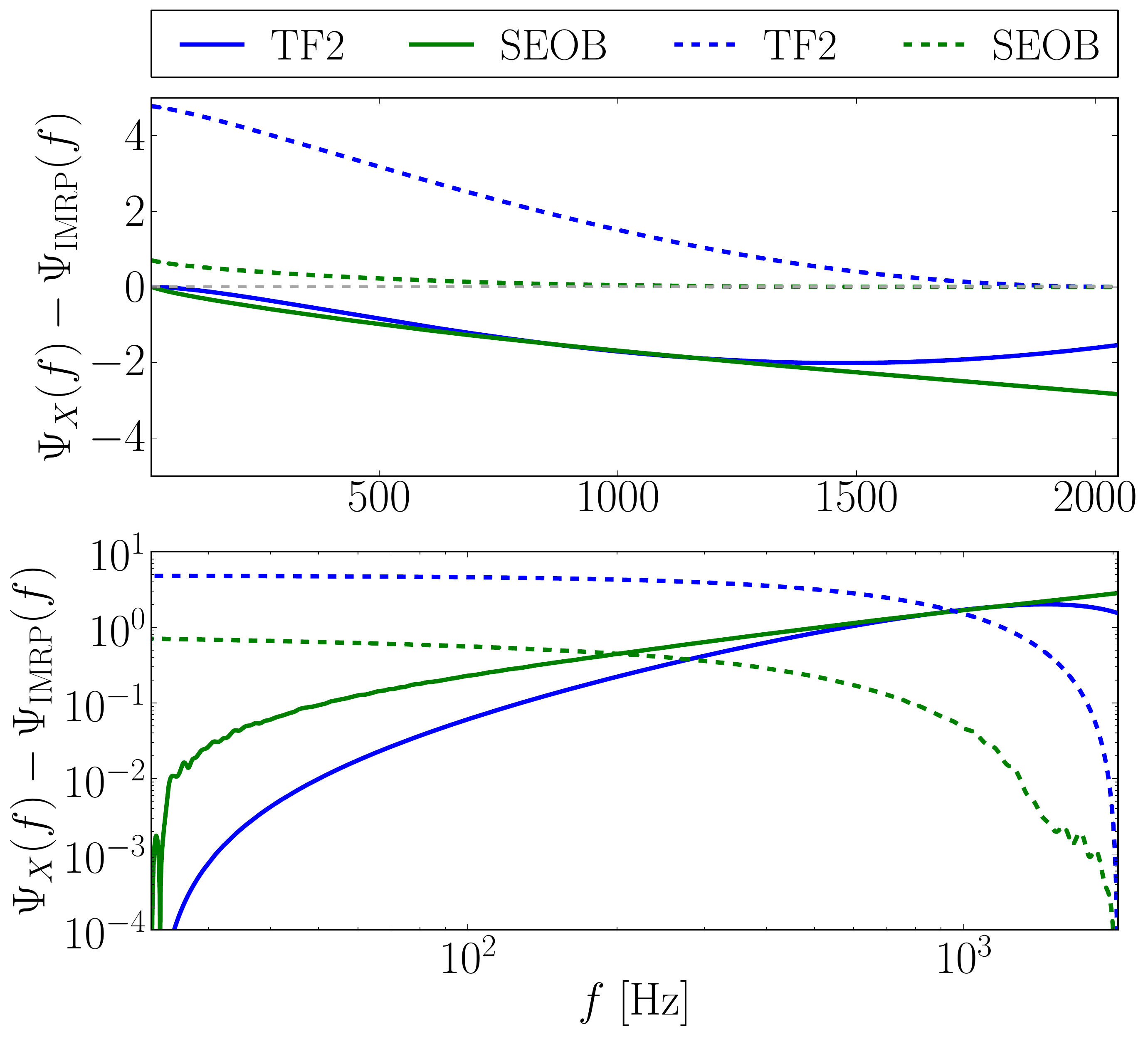}} \quad
   \subfigure{ 
   \includegraphics[keepaspectratio,width=0.48\textwidth]{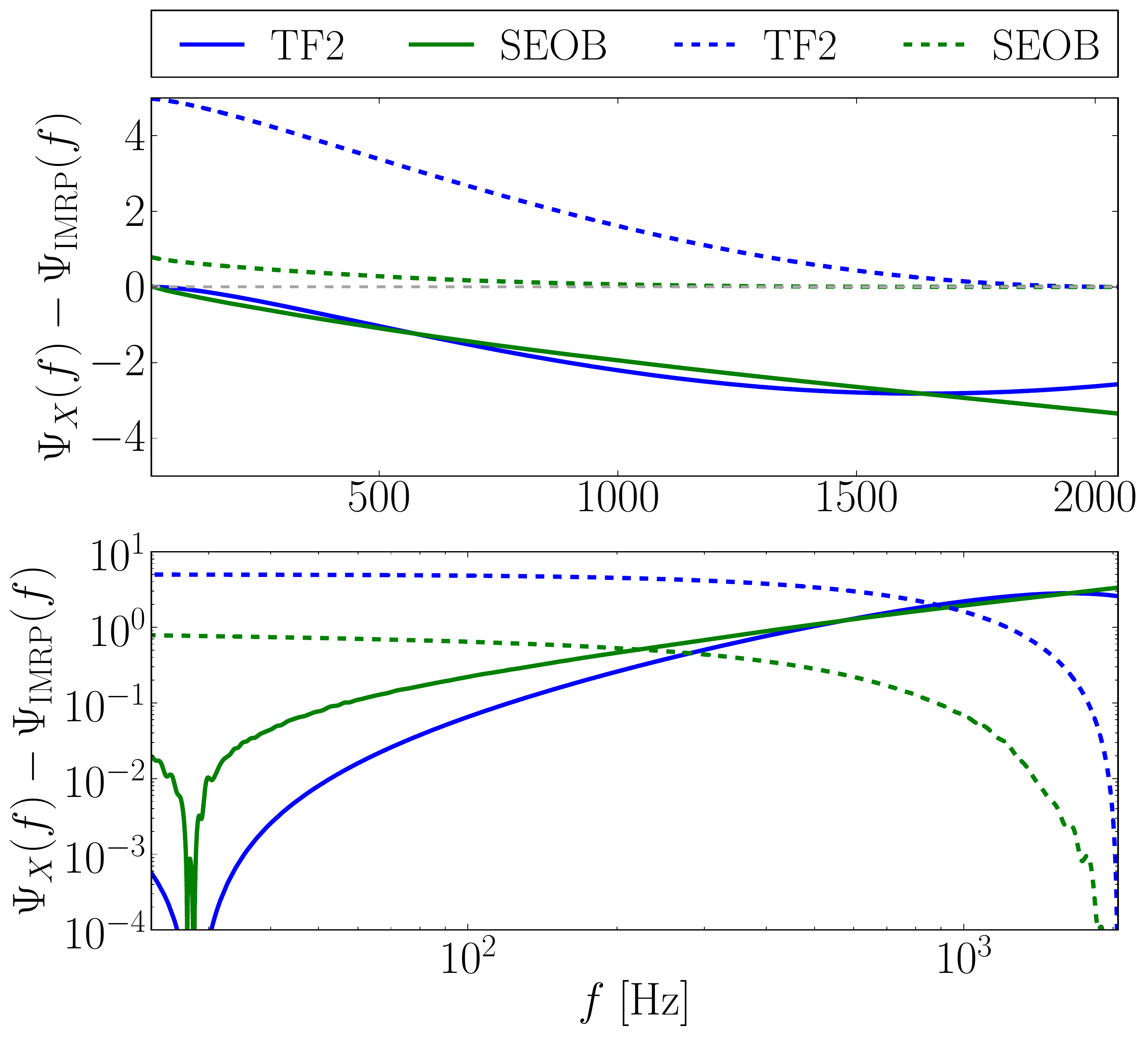}}}
  \caption{Phase difference between \tf\ and \seob\ with respect to \PhenomD for the equal-mass setup (left) 
  and the unequal-mass setup (right). We align the waveforms at $f_0=23\ \rm Hz$ (solid lines) and 
  $f_0=2048~\rm Hz$ (dashed lines) for the equal-mass setup. In the unequal-mass setup, we align the waveforms at
  $f_0=28\ \rm Hz$ (solid lines) and $f_0=2048\ \rm Hz$ (dashed lines).}
  \label{fig:PsiPP}
\end{figure*} 
As an alternative approach and as an effective representation of tidal 
effects beyond the known analytic knowledge, we write the tidal phase as 
\begin{equation}\label{eq:tD_general}
\Psi_{\rm NRT} = - \kappa_{\rm eff}^T
\frac{c_{\rm Newt} x^{5/2}}{X_A X_B} P^{\rm NRT}_\Psi(x) \ ,
\end{equation}
with 
\begin{align}
\label{eq:Pnrtides}
P^{\rm NRT}_\Psi=\frac{1  + \nt_1 x + \nt_{3/2} x^{3/2} + \nt_2 x^{2} + 
\nt_{5/2} x^{5/2}}{1+ \dt_1 x + \dt_{3/2} x^{3/2}}
\end{align}
where $x$ is defined above, $\ct_{\rm Newt} = 39/16$ and $\dt_1 = \nt_1 - 3115 /1248$.
The remaining parameters are determined by fitting and read
$(\nt_1,\nt_{3/2},\nt_{2},\nt_{5/2})=(-17.428,31.867,-26.414,62.362)$
and $\dt_{3/2}=36.089$. We have explicitly set $\dt_1=(\nt_1-\ct_1)$ to ensure the 
recovery of the analytic PN knowledge up to 6PN. 
For details about the calibration procedure, we refer to 
Refs.~\cite{Dietrich:2017aum,Dietrich:2018uni}.
$\kappa_{\rm eff}^T$ is known as the effective tidal coupling constant given by
\begin{equation}
 \label{eq:kappaEff}
 %\kappa_{\rm eff}^T = \frac{2}{13} \left[ \left( 1+12\frac{X_B}{X_A} \right)\left(\frac{X_A}{C_A} \right)^5 k_2^A 
 %+ (A \leftrightarrow B)  \right]
 \kappa_{\rm eff}^T = \frac{3}{13} \left[ X_A^4 \Lambda_A \left( 12 - 11 X_A \right) + (A \leftrightarrow B)  \right]
\end{equation}
When using waveform models with the \texttt{Phenom} and \seob\ PP phasing, 
we use the termination frequency
determined by the tidal coupling constant $\kappa_2^T$ given by
\begin{equation}
f_{\mathrm{merger}} = \frac{1}{2\pi} \omega_0 \sqrt{\frac{X_B}{X_A}} \frac{1+n_1\kappa_2^T+n_2(\kappa_2^T)^2}{1+d_1\kappa_2^T+d_2(\kappa_2^T)^2},
 \label{eqn:fmerger}
\end{equation}
with 
\begin{equation}
\kappa_2^T = 3 X_A X_B \left[ X_A^3 \Lambda_A + (A \leftrightarrow B) \right],
 \label{eqn:kappaT} 
 \end{equation}
where $n_1=3.354\times10^{-2}$, $n_2=4.315\times10^{-5}$, $d_1=7.542\times10^{-2}$ and $d_2=2.236\times10^{-4}$. 
The parameter $\omega_0=0.356$ is chosen such that for 
$M_A=M_B$ and $\kappa_2^T \rightarrow 0$, the non-spinning BBH limit is 
recovered~\cite{Dietrich:2018uni}. The coupling constant $\kappa_2^T$ is closely related to $\kappa_{\rm eff}^T$
and it is indeed the latter which is used in \texttt{LALSuite}.

In addition to the discussed tidal effects which are independent of the NS' spin, 
spin effects depending on the EOS of the NSs enter already at 2PN. 
These spin induced quadrupole-monopole (QM) effects are included in the \PhenomP up to 3PN. 
The spin-induced quadrupole moment~\cite{Poisson:1997ha} is calculated from the 
tidal parameter of each NS using the quasi-universal relations of~\cite{YAGI20171}. 
While in this article the EOS-dependent QM effects are only included in the \PhenomP
and \tf\ approximants, 
this has purely historical reasons. In fact, a new implementation of the \seobnrt\ approximant 
also includes these effects, but was not available when our injection study was started. 

\subsection{Injection study} 

\begin{table*}
\begin{tabular}{l|c|c|c|c|c|c|c|c}
 & $M_A \ [\rm M_\odot]$ & $M_B \ [\rm M_\odot]$ & $M \ [\rm M_\odot]$ & $q=\frac{M_B}{M_A}$ & $\Lambda_A$ & $\Lambda_B$ & $\tilde{\Lambda}$ & $D_L \ [\mathrm{Mpc}]$ \\
\hline
 Source-1  & 1.375 & 1.375 & 2.75 & 1 & 292 & 292 & 292 & 50  \\
 Source-2 & 1.68 & 1.13 & 2.81 & 0.67 & 102.08 & 840.419 & 292 & 50 \\
\end{tabular}
\caption{Overview about the injected sources. 
The columns refer to the individual masses, the total mass, the
mass ratio, the individual tidal deformabilities, the effective tidal deformability $\tilde\Lambda$ and the 
luminosity distance to the source.
Both sources are non-spinning.}
\label{tab:params}
\end{table*}

We perform simulations with sources in agreement with the EOS constraints obtained 
from Refs.~\cite{Abbott:2018exr,Abbott:2018wiz}.
All simulations are performed in stationary Gaussian noise assuming 
LIGO and Virgo design sensitivity~\cite{Aasi:2013wya,AdvLIGOPSD}. For the first source, we choose an equal-mass
binary of component mass 1.375 $\mathrm{M}_\odot$ and component tidal deformability of 
$\Lambda_A=\Lambda_B=292$.
To investigate further the effect of mass-ratio on the estimates of tidal deformability, we 
perform a second set of simulations with $M_A= 1.68, M_B = 1.13$ and  
$\Lambda_A=102.08, \Lambda_B=840.419$, cf.~Tab.~\ref{tab:params}.

Both sources are at a luminosity distance of $50\ \rm Mpc$ and placed face-on. 
The right ascension and declination are $60$ degrees each. 
All considered sources are non-spinning. 
The effects of spin and precession need 
an extended set of injection setups and, therefore, a 
large amount of additional computational resources. 
Thus, we postpone this analysis to the future. 
The masses and tidal deformabilities of the sources are 
summarized in Tab.~\ref{tab:params}. The inference is performed with a 
Bayesian approach.

\subsubsection{Bayesian inference}
We follow a Bayesian approach for parameter estimation, using the built-in 
inference module \linf~\cite{Veitch:2014wba} in the \texttt{LALSuite} package. 
The sampling is done with the Markov Chain Monte Carlo (MCMC) algorithm within 
\linf, called \texttt{lalinference\_mcmc}~\cite{Rodriguez:2013oaa}. 
In the Bayesian framework, all information about the parameters 
of interest is encoded in the posterior probability distribution function (PDF) given 
by Bayes' theorem:
\begin{equation}
 p(\vec{\theta}|\mathcal{H}_s,d) = \frac{p(\vec{\theta}|\mathcal{H}_s)p(d|\vec{\theta},\mathcal{H}_s)}{p(d|\mathcal{H}_s)},
 \label{eqn:Bayes}
\end{equation}
where $\vec{\theta}$ represents the parameter set and $\mathcal{H}_s$ is the hypothesis that a 
GW signal depending on the parameters 
$\vec{\theta}$ is present in the data $d$.
In our setting, the parameter set consists of the parameters common to a BBH 
signal $\{ M_A, M_B, \chi_A, \chi_B, \alpha, \delta, \iota, \psi, D_L, t_c, \varphi_c \}$, and 
in addition the two tidal deformability parameters $\Lambda_A$ and $\Lambda_B$ characteristic to 
a BNS system.
Henceforth, we will follow the mass-weighted tidal deformability 
introduced by, e.g.~\cite{Flanagan:2007ix}
\begin{equation}
\tilde{\Lambda}=\frac{16}{13} \sum_{i=A,B} \Lambda_i \frac{m_i^4}{M}\left( 12-11\frac{m_i}{M} \right), 
\end{equation}
since it is the best determined tidal deformability parameter, where
$M=M_A+M_B$ is the total mass of the binary with component masses
$M_A$ and $M_B$. $\chi_i$ is the dimensionless 
spin parameter given by $\chi_i = \frac{|\vec{S}_i|}{M_i^2}$. $t_c$ and $\varphi_c$ are the 
time at coalescence and reference phase at that instant, respectively. 
We shall use the effective spin parameter defined as a linear combination of the component spins as
$\chi_{\mathrm{eff}} = \frac{M_A \chi_A + M_B \chi_B}{M}$.
%describing the spin aligned with the direction of the 
%orbital angular momentum and the parameter $\chi_\mathrm{p}$, describing the 
%in-plane precessing spin.
\tf\ and \seob\ waveforms are sampled in component spins $\chi_i$ but 
the effective-spin parameter is computed using the above and reported in the 
following analysis.
$D_L$ is the luminosity distance to
the source and $\alpha$ and $\delta$ respectively denote the angles of right ascension and declination.
$\iota$ and $\psi$ are the inclination angle and the polarization angle respectively, describing the binary's
orientation with respect to the detector.
Assuming the noise to be 
Gaussian, the likelihood of obtaining a signal $h(t)$ in data $d$ is given by the proportionality
\begin{equation}
 p(d|\vec{\theta},\mathcal{H}_s) \propto \exp{\left [-\frac{1}{2}(d-h|d-h)\right ]}.
 \label{eqn:lhood}
\end{equation}
In the presence of a GW signal, the data stream output from the detector is 
\begin{equation}
 d = h(t) + n(t),
\end{equation}
where $h(t)$ is the GW signal and $n(t)$ is the noise. 
The scalar product between two functions $a,b$ is 
\begin{equation}
(a|b) = 4 \ \mathrm{Re}\int_{f_{\mathrm{low}}}^{f_{\mathrm{high}}} df\,\frac{a(f) b^*(f)}{S(f)}.
\end{equation}
$S(f)$ refers to the power spectral density (PSD) of the detector. The signal-to-noise ratio (SNR)
of the signal $h$ is defined as
\begin{equation}
 \mathrm{SNR}^2 = 4 \int_{f_{low}}^{f_{high}} df \frac{|\tilde{ h(f)}|^2}{S(f)} ,
\end{equation}
where $\tilde{ h(f)}$ is the signal in the frequency domain.

All priors are motivated by the study of GW170817, Ref.~\cite{Abbott:2018wiz}.
Consequently, the recovery is done with a uniform prior on 
the component tidal deformabilities $\Lambda_A$ and $\Lambda_B$ between 0 and
5000. We sample on distance uniform in co-moving volume up to 100 Mpc. 
Priors on dimensionless spin magnitudes are distributed uniformly between 0 and 0.05. 
The chirp mass $\mathcal{M}=(M_A M_B)^{3/5}/M^{1/5}$, 
is sampled uniformly between 1.184 and 2.168 with 
mass-ratio $q=M_B/M_A$ restricted between 0.125 and 1. 
The sky position as well
as the inclination of the binary are 
uniformly distributed on the sphere.

\section{The imprint of the point-particle baseline}
\label{sec:results:PP}
\subsection{Theoretical modeling}

\begin{figure*}[t]
\includegraphics[keepaspectratio,width=\textwidth]{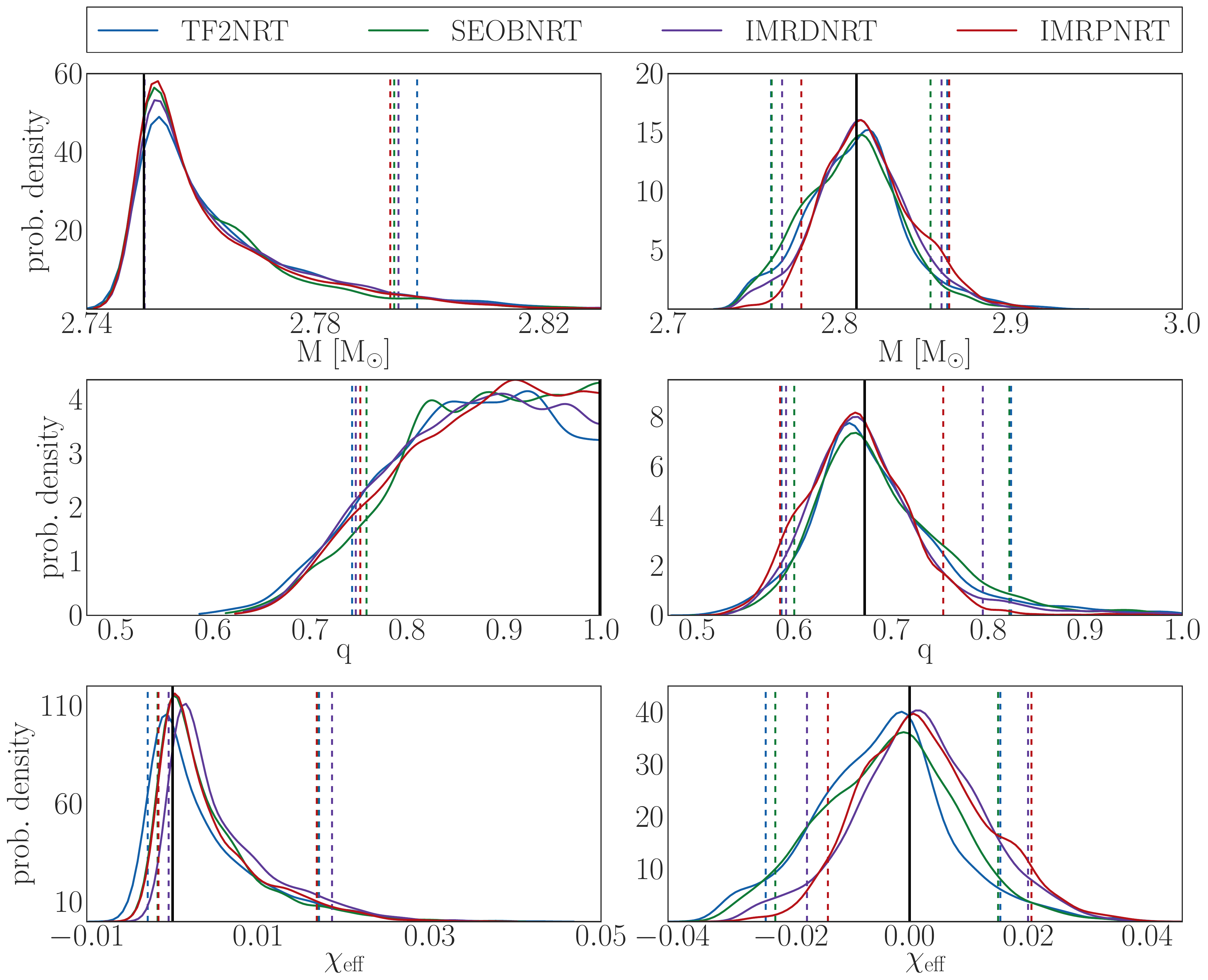} 
  \caption{Posterior PDFs of total mass (top row), mass ratio (middle row) and the effective spin parameter 
          $\chi_\mathrm{eff}$ (bottom row) for the 4 PP models, the tidal phasing being given for all by the \NRtidal
          model. The sources injected are an equal-mass injection (left panel) 
          and an unequal-mass injection (right panel). 
          The vertical dashed lines mark the 5\% and 95\% quantiles, enclosing a 90\% credible interval. 
          Note that for the equal mass case the lower bound of the 90\% confidence interval is indistinguishable from 
          the injected value (black vertical line).}
  \label{fig:mtot_and_chiEff_NRTs}
\end{figure*} 

\begin{figure*}[t]
   \includegraphics[keepaspectratio,width=\textwidth]{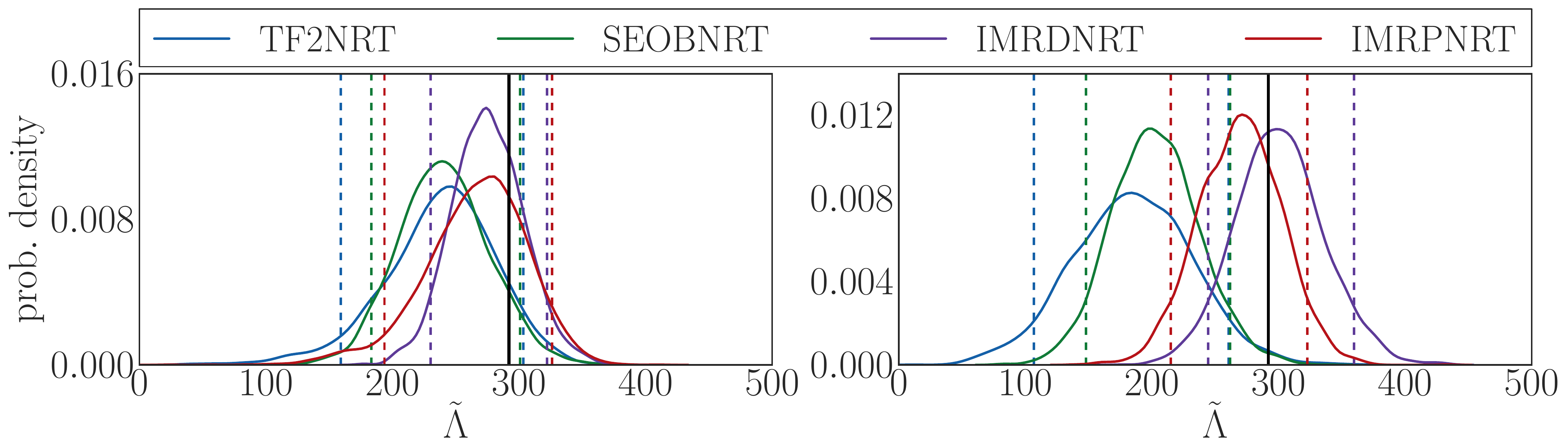}
  \caption{Recovery of tidal deformability $\tilde{\Lambda}$ from a simulation of
          a non-spinning equal-mass binary of component masses 1.375 $\Msun$ each and 
          a tidal deformability $\tilde{\Lambda}$ of 292 (left figure) and the same 
          for an unequal-mass source of 1.68 + 1.13 $\Msun$ and $\tilde{\Lambda}$ of 292 (right figure).
          The simulated waveform model is \imrpnrt, the tidal descriptions
          in the simulated and recovery models are the \NRtidal model. 
          The recovery models are 
          \imrpnrt\ (in red), \tfnrt\ (in blue), \imrdnrt\ (in purple) and \seobnrt\ (in green).
          The vertical lines correspond to the 5\% and 95\% quantiles respectively.
          The injected values are shown as a vertical black line.}
  \label{fig:NRTRecs}
\end{figure*} 

Within this work, we compare three different types of aligned-spinning PP
baseline models: the analytical \tf\ model, 
the EOB-based \seob\ model, and the phenomenological \PhenomD approximant. 
The \PhenomP baseline is obtained by `twisting up' \PhenomD. 
As we inject a non-spinning \imrpnrt\ signal, we shall be referring 
in the following to the \imrdnrt\ waveform, which in the non-spinning 
limit has the same PP baseline.

The phase differences between the waveform models with respect to 
the \PhenomD approximant are shown in Fig.~\ref{fig:PsiPP} 
on a linear scale (top panel) and 
logarithmic scale (bottom panel). The left column refers to the equal-mass, 
the right column to the unequal-mass setup.
To compute the phase difference, we align 
the individual waveforms at a frequency of $f_0=23~\rm Hz$ for the equal-mass 
setup, whereas for the unequal-mass setup, we align the individual waveforms 
at a frequency $f_0=28~\rm Hz$ as the oscillatory behavior for \seob 
(caused by the Fourier transformation and ROM computation) makes the 
alignment non-trivial. During the alignment procedure, 
we set the phase $\psi(f_0)$ and the first derivative $\partial_f \psi|_{f_0}$
of the waveform approximants to be equal at $f_0$. 

We find that the 
phase differences between \tf/\seob\ and \PhenomD are comparable 
and of the order of about 2-3 radians for the equal-mass case 
and about 1 radian larger for the unequal mass case. 
Interestingly, both the phase differences are negative. 
This indicates that \PhenomD is less attractive than the two other approximants. 
Consequently, one could expect that the more attractive point particle models 
\tf\ and \seob\ will predict smaller tidal phase corrections 
than the \PhenomD model under the assumption that 
all the other recovered parameters are identical. 

For completeness, and also because of difficulties in aligning the 
\seob\ waveform due to oscillatory behavior of the 
waveform's phase at low frequencies,
we tested an alignment for $f_0=2048~\rm Hz$, i.e., 
the end of the considered frequency window. For this case the dephasing between 
\seob\ and \PhenomD is less than 1 radian and for \tf\ and \PhenomD is about 5 radians, 
with slightly larger phase difference for the unequal mass setup. 
This clearly shows that the \seob\ and \PhenomD approximant 
are more similar during the late 
inspiral, close to the moment of 
merger, than \tf\ and \seob.

\subsection{Estimating non-tidal parameters}
 \label{subsec:pp_pe_nontidal}
We study the impact of the PP baseline on 
the estimation of the intrinsic parameters by adding the 
\NRtidal contribution to the 4 PP baselines described in 
Tab.~\ref{tab:waveform_models}. 
Figure~\ref{fig:mtot_and_chiEff_NRTs} shows the recovery of the intrinsic parameters 
(from top to bottom):
total mass $M$, the mass ratio $q$, and the effective spin $\chi_\mathrm{eff}$. 
The results are presented for both the equal-mass source 
(left panel) and the unequal-mass source (right panel), cf.~Tab.~\ref{tab:params}.
We find for both cases that all values are recovered in a way that the 
injected value (solid black line) lies within the 90\% credible interval (dashed lines). 
Considering the recovery of the total mass, one finds that the equal mass 
setup shows more accurate recovery visible by the smaller credible interval; 
note the different abscissa ranges of the top panels. 
However, for the equal-mass source, the injected total mass lies at the edge of the 5\% quantile. 
This is because of the condition of $M_A \ge M_B$ imposed during sampling and
is further evidenced in the upper bound on the posterior PDF for $q$ for the same source.
For the unequal mass system, we find that the total 
mass is well constrained, being {$M=2.81^{2.86}_{2.77}$} for 
the \imrdnrt\ recovery, with the 
estimates from the other approximants being almost identical.

Considering the recovery of the mass-ratio $q$, for the equal-mass case, 
we report the lower bound at 90\% confidence and show similar recovery for 
the different waveform approximants. For the unequal-mass case however, 
the recovered value peaks around the injected one, and is also 
consistent among all the models. This emphasizes
the small systematic bias with respect to different waveform approximants. 

The recovery of the effective aligned-spin parameter $\chi_\mathrm{eff}$ is similar 
for all the baseline models, with the injected value lying well
within the 5\% and 95\% quantiles in the unequal-mass simulation. 

\subsection{Estimating tidal parameters}
\label{subsec:pp_pe_tidal}
Figure~\ref{fig:NRTRecs} shows the recovery of the tidal 
deformability parameter.
For the equal-mass case, we note that the injected value of $\tilde{\Lambda}$ 
(shown by the black vertical line) is always contained within the 5\% and 95\% quantiles of the posterior 
distribution. The recovery with the \imrdnrt\ model shows the smallest 
offset and spread around the injected value. 
This is caused by the fewer number of parameters of the \imrdnrt\ approximant compared to the 
\imrpnrt\ model, which is precessing. 
This observation supports the use of
spin-aligned models for the extraction of parameters in addition to precessing models. 
The \seobnrt\ and \tfnrt\ model differ from the injected value because of 
the different underlying spin-aligned baseline family.
As suggested by the phasing in Fig.~\ref{fig:PsiPP}, 
we find that \seobnrt\ and \tfnrt\ 
predict smaller tidal deformability due to the 
slightly more attractive PP baseline. 
For the unequal-mass binary, the injected value is 
contained within the 5\% and 95\% quantiles only when the recovery PP baseline 
is the \texttt{Phenom} model. Although the 
other instrinsic parameters show very similar results from 
recovering with different PP baseline models, we note that 
the recovery of the $\tilde{\Lambda}$ parameter 
differs between waveform models with different PP baselines even when the tidal 
phasing is unchanged.
\begin{figure*}[t]
\mbox{\subfigure{
   \includegraphics[keepaspectratio,width=0.48\textwidth]{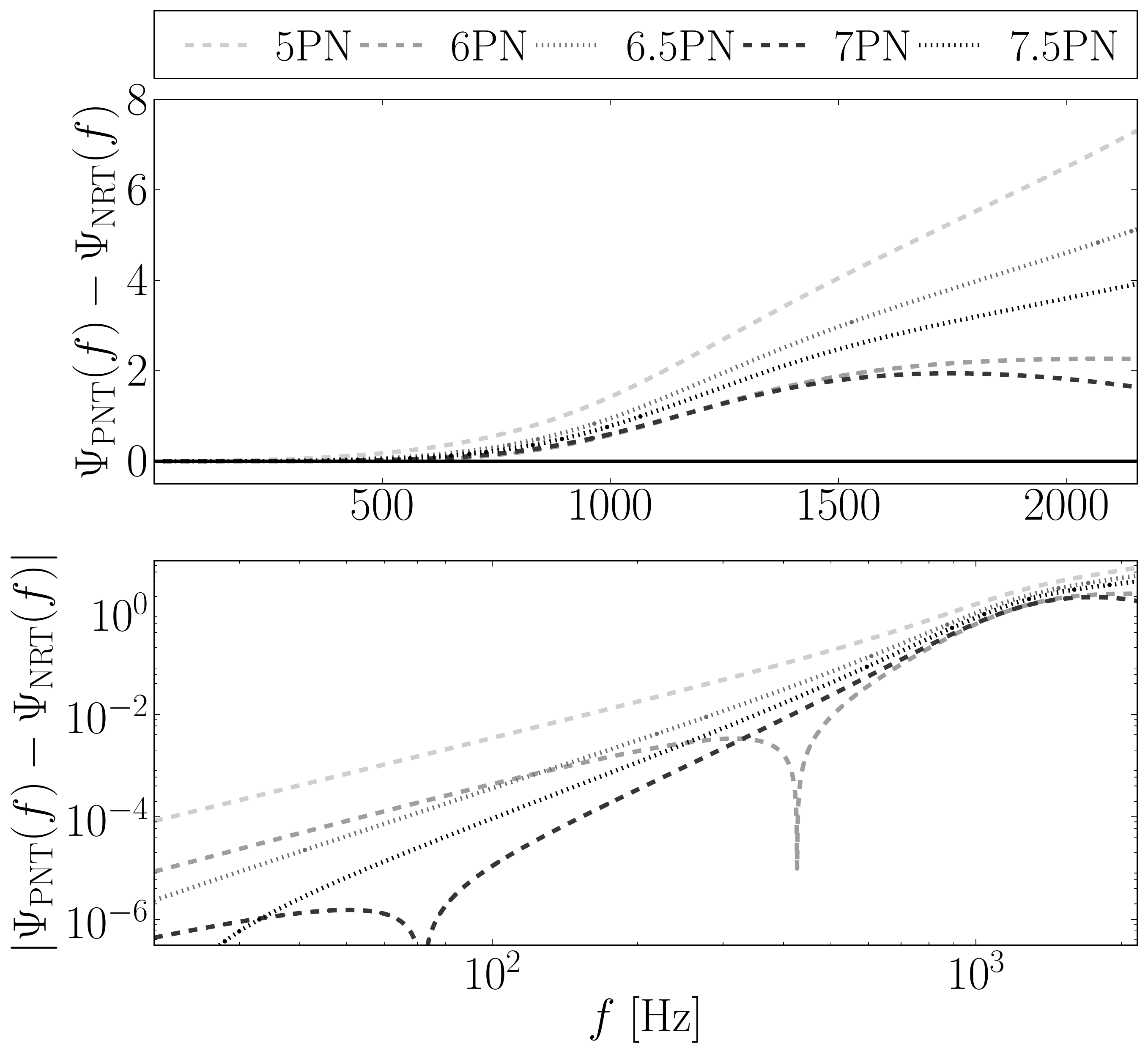}}\quad
   \subfigure{
   \includegraphics[keepaspectratio,width=0.48\textwidth]{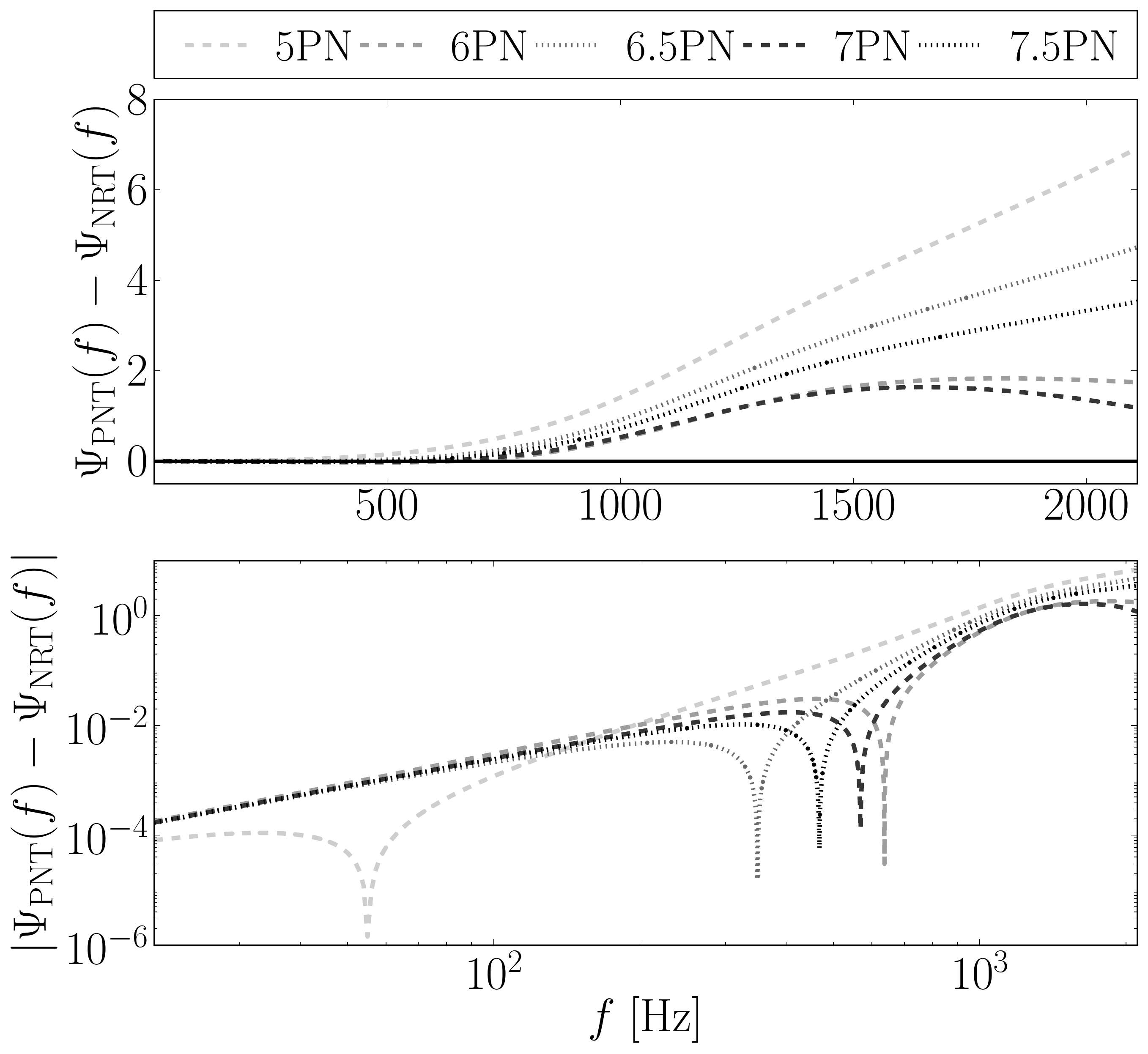}}}
  \caption{Figure showing the difference between the tidal phasing terms from PN approximation
          and the \NRtidal description for an equal-mass BNS source (left figure) and for an unequal-mass 
          BNS source (right figure), cf.~Tab.~\ref{tab:params}. The top panel shows the differences starting 
          from the leading order 
          PN-tidal phase at 5PN through the highest known order at 7.5PN. The bottom panels 
          show the absolute difference of the phasing terms.}
  \label{fig:Psitides}
\end{figure*}
\section{Imprint of the tidal description}
\label{sec:results:tides}

\subsection{Theoretical modeling}
\begin{figure*}[t]
\includegraphics[keepaspectratio,width=\textwidth]{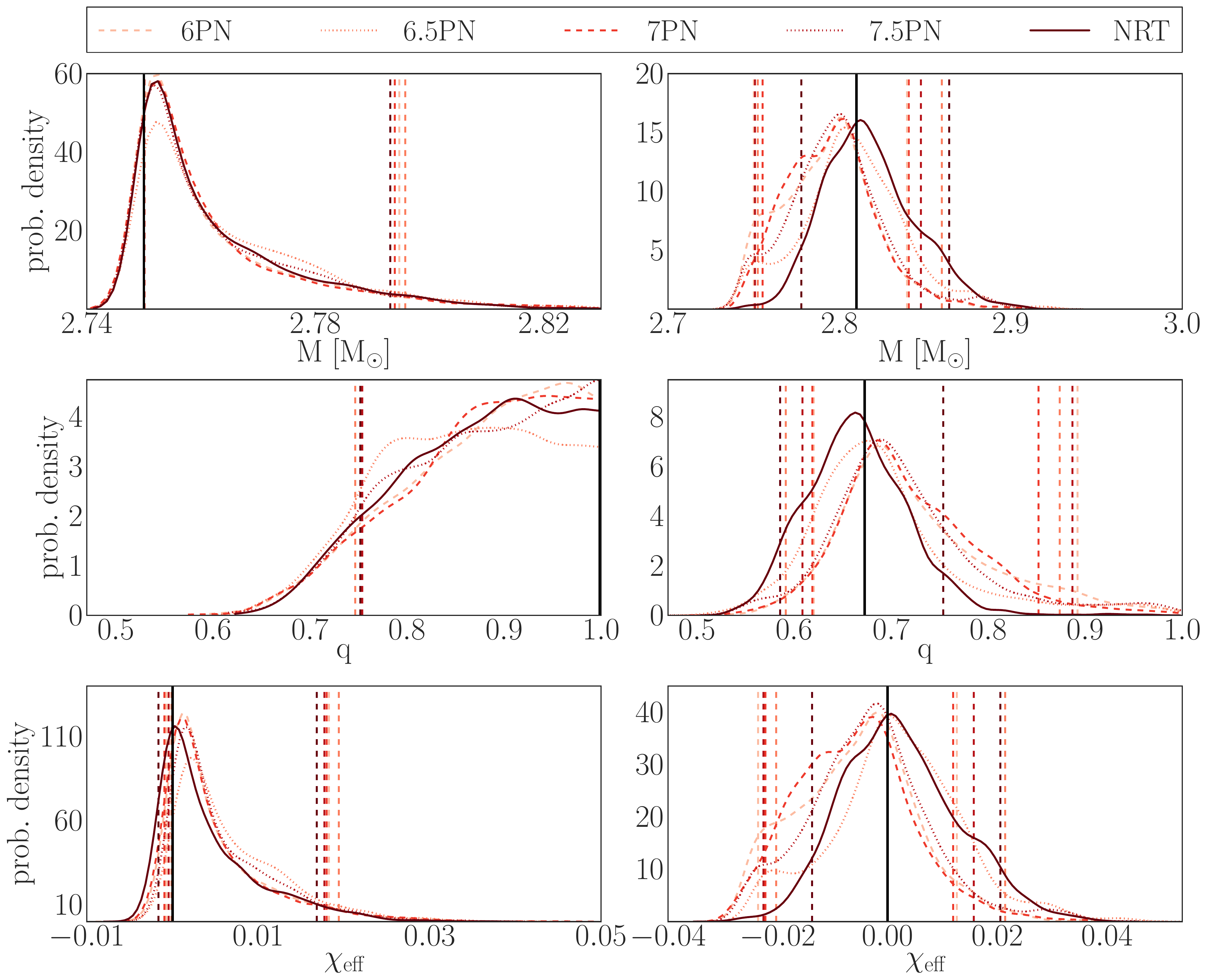} 
  \caption{Posterior PDFs of total mass (top row), mass ratio (middle row) and the effective spin parameter 
          $\chi_\mathrm{eff}$ (bottom row) for the \PhenomP PP model and the tidal phasing given by the 
          PN description starting from 6PN through 7.5PN and the phasing described by the \NRtidal model. 
          The sources injected are an equal-mass binary (left panel) and an unequal-mass binary (right panel),
           cf.~Tab.~\ref{tab:params} The injected values are shown by black, vertical lines whereas the dashed lines show 
           the 90\% credible interval.}
  \label{fig:mtot_and_chiEff_all_tides}
\end{figure*} 

\begin{figure*}[t]
   \includegraphics[keepaspectratio,width=\textwidth]{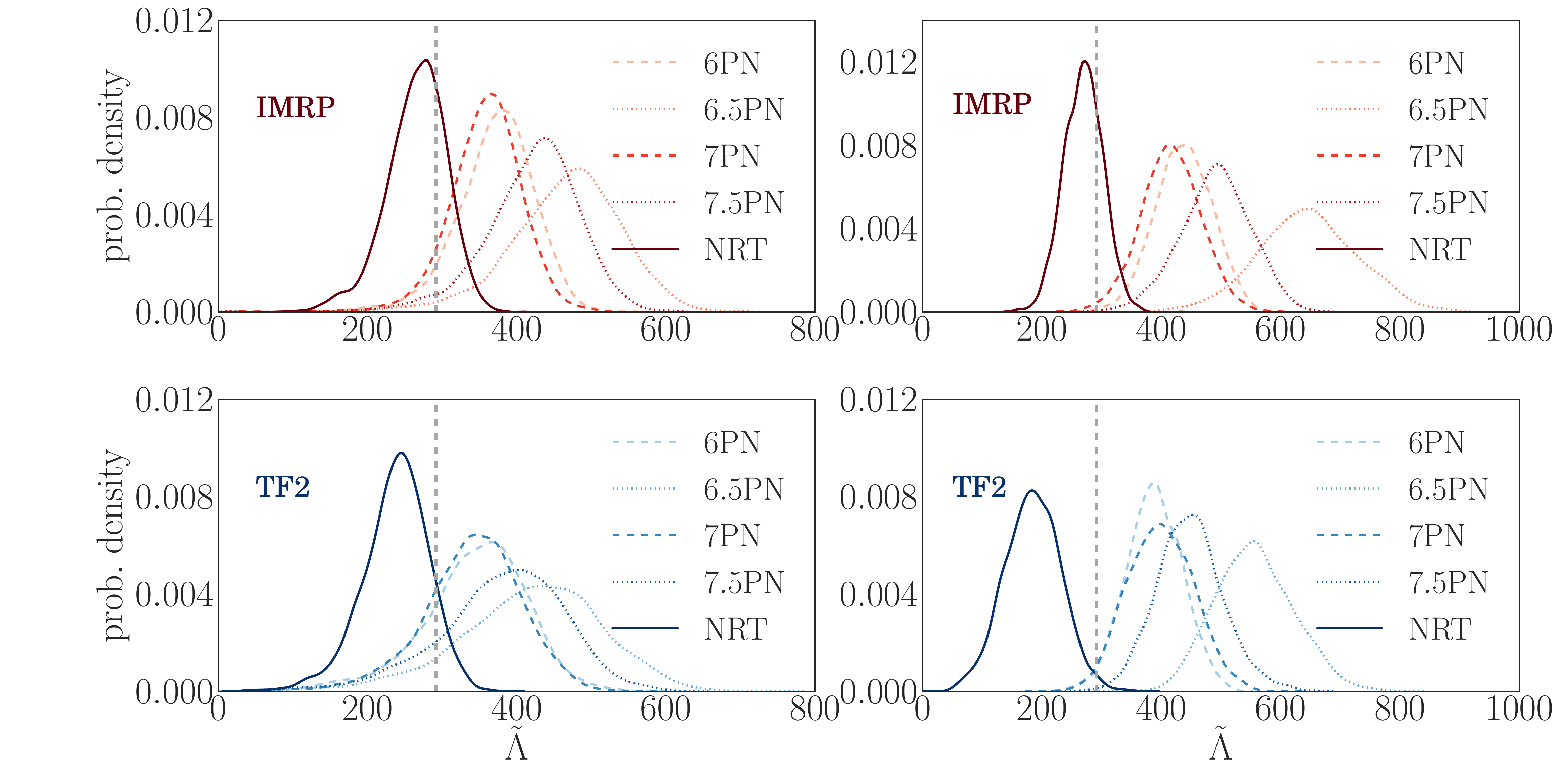}
  \caption{Posterior probability distributions showing estimates of $\tilde{\Lambda}$ from a simulation of 
           of an equal-mass BNS source with \imrpnrt\ (left figure) and an unequal-mass 
           BNS source with \imrpnrt\ (right figure). The point-particle baseline recovery waveform models
           are \PhenomP (top panel) and \tf\ (bottom panel). The tidal 
           phasing in each model is varied from 6PN through 7.5PN and the \NRtidal description.
          The gray dashed line shows the injected value.}
  \label{fig:Rec_all_tidal_models}
\end{figure*} 

\begin{table*}
\begin{tabular}{c| c| c | c | c|c }
\hline
  \multicolumn{2}{c|}{Recovery model} & \multicolumn{2}{c}{Equal-mass case ($\tilde{\Lambda} = 292$)} & \multicolumn{2}{|c}{Unequal-mass case ($\tilde{\Lambda} = 292$)} \\ \hline
  Baseline & Tidal phasing & Estimate of $\tilde{\Lambda}$  & Systematic bias (stacc) &Estimate of $\tilde{\Lambda}$  & Systematic bias (stacc)  \\
\hline
 \tf\  & 6PN & $350.3_{215.8}^{445.1}$ & 87.8 & $388.4_{310.7}^{465.4}$ & 107.0 \\
           & 6.5PN & $425.7_{245.7}^{564.6}$ & 158.3 & $554.1_{452.7}^{669}$ & 272.6 \\
           & 7PN & $343.7_{222.2}^{441}$ & 82.1 & $398.6_{310.7}^{489.8}$ & 120.7 \\
           & 7.5PN & $388.8_{225.7}^{500.8}$ & 122.2 & $446.7_{357.5}^{543.7}$ & 165.6 \\
           & \NRtidal & $241.3_{159.2}^{303.3}$ & 70.4 & $184.6_{106.6}^{260.6}$ & 117.5 \\ \hline
  \PhenomP & 6PN & $375.4_{281.8}^{448.3}$ & 95.2 & $437.5_{357.5}^{514.9}$ & 152.7  \\
           & 6.5PN & $472.4_{333.2}^{578.8}$ & 189.3 & $644.1_{508.0}^{782.9}$ & 362.8 \\
           & 7PN & $360.9_{278.2}^{432.4}$ & 82.6 & $415.9_{331.8}^{498.7}$ & 133.9 \\
           & 7.5PN & $428.0_{308.2}^{512.9}$ & 144.2 & $493.1_{388.3}^{584.8}$ & 207.5  \\
           & \NRtidal & $268.9_{193.6}^{326}$ & 48.8 & $270.0_{214.8}^{322.9}$ & 39.9 \\ \hline
   \PhenomD & \NRtidal & $274.7_{230.2}^{322.1}$ & 32.8 & $299.8_{244.5}^{359.8}$ & 36.3 \\ \hline
   \seob\ & \NRtidal & $239.9_{183.2}^{300.9}$ & 62.5 & $201.7_{147.8}^{261.7}$ & 95.8 \\ \hline
   
\end{tabular}
\caption{Results from parameter estimation study with simulation of 
\imrpnrt\ for an equal-mass and unequal-mass source.}
\label{tab:inj_results}
\end{table*}
As given explicitly in Eq.~\eqref{eq:PsiPNT} there 
exist analytical (although incomplete) PN knowledge up 7.5PN oder. 
To understand the imprint of the employed PN order, we present the phase difference 
between the different PN-tidal phase contributions with respect to the 
\NRtidal model, Eq.~\eqref{eq:Pnrtides}, in Fig.~\ref{fig:Psitides}. 
We employ the \NRtidal model as our benchmark since it has been shown to reliably agree with 
NR waveforms~\cite{Dietrich:2017aum,Dietrich:2018upm} as well as with hybrid 
\TEOBResumS~\cite{Bernuzzi:2014owa,Nagar:2018zoe}/NR hybrids~\cite{Dietrich:2018uni}.
Nevertheless, we stress that also the \NRtidal phasing potentially
overestimates tidal effects with respect 
to TEOBResumS/NR-hybrids~\cite{Dietrich:2018uni,Dudi:2018jzn}. 

The PN-based phasing
terms are considered starting from the single leading-order term at 5PN to the 
higher order terms, which also include the preceding orders. The absolute value of the 
difference of these terms is plotted in the bottom panel of 
the figure on a double logarithmic scale. We note that
the terms at 7PN and 6PN are closer to the \NRtidal description than the half PN orders at 6.5PN 
and 7.5PN. This is related to the half-PN orders being repulsive.
The plot refers to our fiducial equal-mass 
(left) and unequal mass (right) BNS, cf.~Tab.~\ref{tab:params}.

\subsection{Estimating non-tidal parameters}
Figure~\ref{fig:mtot_and_chiEff_all_tides} shows the results from parameter estimation of 
the equal-mass setup (left panel) and the unequal-mass setup (right panel). 
We recover the parameters with \imrpnrt\ and \imrppnt\ 
(with the tidal phasing varying from 
6PN through 7.5PN orders). 
We find overall that for the equal mass configuration the estimation 
of the total mass, mass ratio, and effective spin 
is almost unaffected by the choice of the tidal contribution. 
A similar statement is true when one uses \tf\ as a PP baseline. 
Although the recovered values 
differ more for the unequal-mass configuration,
the injected values are contained within the 5\% and 95\% quantiles
for each approximant. However, the most notable difference 
is that the PN-tidal approximants predict mass 
ratios (credible intervals) larger than the \NRtidal description.

\subsection{Estimating tidal parameters}
Figure~\ref{fig:Rec_all_tidal_models} shows the results considering the 
parameter estimation of the tidal deformability for the 
equal-mass (left panel) and unequal-mass (right panel) binary sources. 
We recover the parameters with the 
\imrpnrt\ and \imrppnt\ models (red) and with 
the \tfnrt\ and \tfpnt\ (blue). 
It is expected that 
the recovery with \imrpnrt\ would be exact, 
except limitations from statistical uncertainties. 
The simulated signal however has a very high SNR, $\sim \ 87$, and we therefore 
ignore presence of statistical biases in our study. 
%Nevertheless, there is a small offset in the 
%deformability parameter $\tilde{\Lambda}$ 
%in the top panel of Fig.~\ref{fig:Rec_all_tidal_models} 
%when using \imrpnrt\ in recovery. 
%This effect is the same as what we observed in Sec.~\ref{subsec:pp_pe_tidal} 
%and is attributed to the fact that in case of \imrpnrt,
%the approximant also allows for the inclusion of precession, 
%increasing the parameter space dimensionality by the presence of the 
%additional effective precessing-spin parameter.

For the equal-mass injection in Fig.~\ref{fig:Rec_all_tidal_models}, as seen also from 
the tidal phasing from Fig.~\ref{fig:Psitides}, 
we find that the recoveries with the 6PN and 7PN orders in \imrppnt\ lie closer to the injected 
\imrpnrt\ approximant than the half-PN orders. We observe a similar trend in the \tf\
recoveries, albeit with a constant shift due to the difference in the PP baseline models between 
the injection and recovery. Table~\ref{tab:inj_results} quotes the median values of $\tilde{\Lambda}$ posterior
PDFs and the 5\% and 95\% quantiles. We note that the injected value is always within this interval.

In case of the unequal-mass injection, the estimates deteriorate and the recovery with \tf\
and \seob\
lead to an incorrect measurement. Table~\ref{tab:inj_results} also summarizes the inferred values of 
$\tilde{\Lambda}$ for the unequal-mass simulation and we find that the injected value lies within the 5\% and 95\% 
quantiles only with the recovery with the \imrpnrt\ and \imrdnrt\ models. We 
quantify the systematic bias with the standard accuracy statistic (stacc) defined as
\begin{equation}
 S = \sqrt{\frac{1}{N} \sum_{i=1}^N (x_i - x_{\mathrm{inj}})^2},
 \label{eq:stacc}
\end{equation}
where $N$ is the number of samples, $x_i$ is the $i^{th}$ posterior sample and $x_{\mathrm{inj}}$ 
is the injected value of the parameter $x$, here $\tilde\Lambda$.
The lowest values of $S$ are indeed obtained for the recovery model \imrdnrt\ and is followed 
by \imrpnrt. When the simulated source is an unequal-mass binary, the statistic is steadily higher than 
for the equal-mass configuration, 
again showing the general deterioration of estimation of the tidal deformability 
parameter.

\section{Conclusion}
\label{sec:conclusion}

We studied possible systematic biases caused by the choice of the 
waveform approximants during the recovery of the source 
parameters from future detections. 
The studied cases had an SNR
of $\sim 87$. Expecting the design sensitivity of 
Advanced LIGO and Virgo to be about three times better 
than the current sensitivity, this SNR would correspond approximately 
to that of a GW170817-like system at design sensitivity.

We find that for such scenarios the existing waveform model
approximants would be sufficient to extract non-tidal parameters 
as masses, mass ratios, and spins. 
On the contrary, the measurement of the tidal deformability can be significantly
biased up to a point where the injected value is not contained within the 
90\% credible interval. 

In particular, the large systematic errors for unequal mass systems, 
might lead to misclassification of BNS, BBH, and neutron star-black hole 
systems. 
Thus, better waveform models with improved tidal descriptions are imperative for 
characterizing unequal-mass BNS and/or NSBH systems in the advanced detector era.
These improvements have to include 
(i) a calibration of the \NRtidal model to unequal mass binaries;
(ii) an incorporation of analytical knowledge beyond 6PN knowledge in the \NRtidal model. 

In a future publication, 
we will address the issue of possible systematics 
arising from the presence of 
spins and precession in the simulated sources and the influence 
of missing spin-tides coupling in the existing waveform approximants. 

\begin{acknowledgments}
  We thank Chris Van Den Broeck, Alessandra Buonanno, Reetika Dudi, 
  Tanja Hinderer, Serguei Ossokine, Jocelyn Read, Ben Lackey, 
  Archisman Ghosh, 
  Jolien Creighton for discussions and helpful comments. 
  In particular, we thank Alessandro Nagar and Sebastiano Bernuzzi 
  for stimulating discussions after GW170817 which partially lead to the presented 
  study and we thank Benji Matthaei for discussions during his 
  Bachelors thesis project in the data analysis group at Nikhef. 
  AS and TD are supported by the research programme 
  of the Netherlands Organisation for Scientific Research (NWO).
  TD acknowledges support by the European Union’s Horizon 
  2020 research and innovation program under grant
  agreement No 749145, BNSmergers. We are grateful for the 
  computing resources provided by
  the LIGO-Caltech Computing Cluster where all our simulations 
  were carried out.
\end{acknowledgments}

% Create the reference section using BibTeX:
\bibliography{paper20181009.bbl}

\end{document}